\documentstyle[preprint,version2,aps]{revtex}
\draft
\begin{document}

\begin{title}
Spin Conductivity and Spin-Charge Separation
in the High $T_c$ Cuprates
\end{title}

\author{Qimiao Si\cite{rice}}
\begin{instit}
Department of Physics, University of Illinois, Urbana, IL 61801
\end{instit}

\begin{abstract}

We study both the spin and electrical conductivities in models of
relevance to the high $T_c$ cuprates. These models describe
metallic states with or without spin-charge separation. We
demonstrate that, given a linear in temperature dependence
of the electrical resistivity, the spin resistivity should also
be linear in temperature in the absence of spin-charge separation
and under conditions appropriate at least for the optimally doped
cuprates, but is in general {\it not} so in the presence of
spin-charge separation. Based on these results, we propose to use the
temperature dependence of the electron spin diffusion constant to
diagnose spin-charge separation in the cuprates.
\end{abstract}

\newpage
%\begin{twocolumn}
%\begin{narrowtext}

One of the most striking behaviors in the normal state of the high $T_c$
cuprates is the linear in temperature dependence of the
electrical resistivity\cite{Batlogg}. The extended temperature
range over which this $T-$linear resistivity occurs, along with other
experimental signatures, have led to the conclusion that the
electrical resistivity is dominated by electron-electron scattering
instead of electron-phonon scattering. In a canonical Fermi liquid,
however, the Pauli exclusion principle strongly suppresses the
phase space available for quasiparticle-quasiparticle scattering,
and the resistivity due to electron-electron scattering is expected
to have a $T^2$ dependence. The observed $T-$linear resistivity,
therefore, suggests that the normal state of the high $T_c$ cuprates
deviates from a canonical Fermi liquid.

The precise nature of the normal state, and exactly how electron-electron
scattering leads to this $T-$linear resistivity, remain subjects of
debate. Starting from the work of Anderson\cite{Anderson}, spin-charge
separation has been suggested to characterize the normal state. In a
spin-charge-separated non-Fermi liquid, the elementary excitations are
divided into two species, each carrying either spin or charge quantum
numbers only. Such a decomposition alters the phase space available
for electron-electron scattering and can, in some cases, result in a
$T-$linear resistivity. Spin-charge separation occurs in the Luttinger
liquid in one dimension\cite{1Dlut} and possibly in two dimensions
as well\cite{Anderson}, in a phase of the two dimensional $t-J$ model
with massless transverse gauge fields\cite{Gauge}, and in a mixed valence
state of an extended Hubbard model in infinite dimensions with competing
spin and charge fluctuations\cite{SK}. Alternatively, it has been proposed
that some form of Fermi liquids with low energy scales can describe the
normal state\cite{Review,Monthoux,mfl}. These states represent minimal
modifications of the canonical Fermi liquid. The elementary excitations
are quasiparticle-like, carrying both spin and charge degrees of freedom;
these quasiparticles are coupled to some soft collective charge and/or
spin fluctuations. While the resistivity is necessarily quadratic in $T$
below the soft energy scale, $T^*$, it can become linear in $T$
at $T \gg T^*$.

While the existence or absence of spin-charge separation remains to be
established, it is worth noting that a difference does appear to exist
between the spin dynamics and charge transport properties. This is most
clearly seen in the optimally doped LaSrCuO in which data are available
over a wide range of temperatures. The Cu-site NMR relaxation rate,
$({1 \over T_1})_{Cu}$, crosses over from a low temperature $T-$linear
dependence to a high temperature $T-$independent behavior\cite{Imai}
while, over the corresponding temperature range, the electrical resistivity
shows essentially no deviation from the $T-$linear behavior\cite{Batlogg}.
This difference, however, can not be used as direct evidence for spin-charge
separation as the correlation functions measured by these two probes can not
be simply related.

In this communication we propose that a comparison between the
temperature dependences of the electrical and spin conductivities
can clarify whether or not spin-charge separation occurs in the high $T_c$
cuprates. To be concrete, we consider several models that might be relevant
to the physics of the metallic cuprates. We find that, given a $T-$linear
electrical resistivity, and under conditions appropriate at least for the
optimally doped cuprates, the spin resistivity is necessarily linear in
temperature in the absence of spin-charge separation, but is in general
{\it not} so in a spin-charge separated state. We note that, the spin
conductivity ($\sigma_s$) describes the response of the spin current
($j_s$) to a gradient of magnetic field\cite{notetensor}. The Einstein
 relation\cite{Forster} states that $\sigma_s = \chi_s D_s$. Here,
$\chi_s$ is the uniform spin susceptibility. $D_s$ is the spin diffusion
constant, which can be measured using the technique of
spin-injection\cite{Hass,Johnson}. It should be noted that the measurement
of spin conductivity is feasible in the cuprates as the effective
interactions induced by the spin-orbit coupling are relatively small
in the cuprates [of the order of a few $meV$\cite{Thio}].

\noindent
\underline{\bf Without Spin-Charge Separation:}
In a Fermi-liquid-like state, a $T-$linear resistivity can arise from
quasiparticles being scattered from soft collective fluctuations.
We can study the conductivities in these states within the following
phenomenological action,

\begin{eqnarray}
{\cal S} = && {\cal S} _{qp} +  {\cal S}_{collective}
+  {\cal S}_{int}\nonumber\\
 {\cal S}_{qp} = &&\int d\omega \sum_{k\sigma} c_{k\sigma}^{\dagger}
( - \omega + \epsilon_k ) c_{k\sigma}\nonumber\\
 {\cal S}_{collective} = &&\int d\omega \sum_{q}
[N_q \chi_{cf}^{-1} (q, \omega ) N_q
+{\vec S}_q \chi_{sf}^{-1} (q, \omega )
\cdot {\vec S}_q]\nonumber\\
 {\cal S}_{int} = &&\int d\omega \sum_{qk}
[V_q (\sum_{\sigma}c_{k+q~ \sigma}^{\dagger} c_{k\sigma}) N_q +
J_q (\sum_{\sigma \sigma'}c_{k+q ~ \sigma}^{\dagger}
{\vec s}_{\sigma \sigma'} c_{k \sigma'} ) \cdot {\vec S}_q ]
\label{lagfl}
\end{eqnarray}
Here, $ {\cal S}_{qp}$ describes the single particle states, created by
$c_{k\sigma}^{\dagger}$, with a dispersion $\epsilon_k$.
$ {\cal S}_{collective}$ describes overdamped collective charge
and spin degrees of freedom, $N_q$ and ${\vec S}_q$,
with fluctuation spectra $\chi_{cf} (q , \omega )$ and
$\chi_{sf} (q , \omega )$, respectively.
Finally, $ {\cal S}_{int}$ describes the coupling of the single particle
states to the collective fluctuations. $V_q$ and $J_q$ are the
charge and spin coupling constants, respectively.
Eq. (\ref{lagfl}) is quite general. It incorporates as special cases
several proposed scenarios for the cuprates. For instance, the marginal
Fermi liquid approach\cite{mfl} corresponds to choosing both
$Im\chi_{sf}(q, \omega )$ and $Im \chi_{cf} (q , \omega )$ to be
$\rho_o{\rm sgn} \omega$ for $\omega>T$, and $\rho_o \omega /T$ for
$\omega < T$ (where $\rho_o$ is the density of states). The nearly
antiferromagnetic Fermi liquid approach\cite{Monthoux} corresponds
to neglecting charge fluctuations, and assuming a mean field form for
$Im\chi_{sf}(q, \omega )$. Other scenarios, as reviewed in
Ref. \cite{Review}, are also incorporated in Eq. (\ref{lagfl})
in a similar fashion.

We calculate the electrical and spin conductivities using the
Kubo formula. The electrical and spin current operators
associated with the quasiparticles
are given by $\vec j=-e \sum_{k\sigma}  {\vec v}_k
c_{k\sigma}^{\dagger} c_{k\sigma}$ and
${\vec j}_s=(g\mu_B/2) \sum_{k\sigma} {\vec v}_k \sigma
c_{k\sigma}^{\dagger} c_{k\sigma}$, respectively,
where ${\vec v}_k = \partial \epsilon_k / \partial {\vec k}$.
The current-current correlation functions
are evaluated to the leading nonvanishing order within
the memory function formalism\cite{Goetz}. This is equivalent
to the semi-classical approach through solving the linearized
Boltzmann equation for the quasiparticle distribution
function\cite{Goetz}. Diagrammatically, it amounts to a resummation
of the conductivity diagrams including the vertex corrections
(ladder diagrams only) and self-energy corrections (to the leading
non-vanishing order). The standard assumption made in this procedure
is that there exists enough electron-electron Umklapp scatterings
so that the scattering of the quasiparticles off of the
collective fluctuations do contribute
to the dissipation of the electrical current. This condition,
while not satisfied for a jellium model, is expected to be well
satisfied when an underlying lattice exists, and when the Fermi
surface is large. The latter is well established
at least for the optimally doped cuprates.
The resulting expressions of the electrical and spin resistivities,
$\rho$ and $\rho_{spin} = 1/ \sigma_s$, can be written in terms of
the transport scattering rate, ${1 \over \tau_{tr}}$, and the spin
transport scattering rate, ${1 \over \tau_{tr,s}}$, respectively:
$\rho = {4 \pi \over \omega_p^2} {1 \over \tau_{tr}}$, and
$\rho_{spin} = {4 \pi \over \omega_{p,s}^2} {1 \over \tau_{tr,s}}$.
Here, ${\omega_p^2 / 4 \pi} = e^2 A$
and ${\omega_{p,s}^2 / 4 \pi} = (g \mu_B/2)^2 A$, where
$A=N (\epsilon_F) <<v_{kx}^2>>_{FS}$,
and ${1 \over \tau_{tr}}$ and ${1 \over \tau_{tr,s}}$
are given by ${1 \over \tau_{tr}} = \left( {1 \over \tau_{tr}}
\right)_{cf} + \left({1 \over \tau_{tr}}\right)_{sf}$ and
${1 \over \tau_{tr,s}} = \left( {1 \over \tau_{tr,s}} \right)_{cf} +
\left( {1 \over \tau_{tr,s}} \right)_{sf} $. Here
$\left ({1 \over \tau_{tr}} \right)_{cf}$,
$\left ( {1 \over \tau_{tr,s}} \right)_{cf}$,
and $\left ( {1 \over \tau_{tr}} \right)_{sf}$, $\left (
{1 \over \tau_{tr,s}} \right)_{sf}$ correspond to
the contributions from electron scatterings off of charge
fluctuations and spin fluctuations respectively.

Consider first the contribution from charge fluctuations alone.
We find that $\left( {1 \over \tau_{tr}} \right)_{cf}$ and $\left( {1 \over
\tau_{tr,s}}\right)_{cf} $ are equal and given by

\begin{eqnarray}
\left( {1 \over \tau_{tr}} \right) _{cf} = \left( {1 \over \tau_{tr,s}}
\right)_{cf} = {1 \over  A}
\sum_{k, q} \gamma_{kq}^2 B(k, q) V_q^2
Im \chi _{cf} (q, \epsilon_{k+q}-\epsilon_{k})
\label{taucf}
\end{eqnarray}
where $\gamma_{kq} = v_{k+q} - v_k$ is the difference between the group
velocities of the quasiparticles before and after a scattering event, and
$B(k,q) = (-\partial n_b(\epsilon)/\partial \epsilon)_{
\epsilon_{k+q} - \epsilon_k } [f(\epsilon_{k+q})-f( \epsilon_k )]$
with $n_b(\epsilon)$ and $f(\epsilon)$ representing the boson and
fermion distribution functions. The scattering
rates become linear in $T$ when the integrated spectral weight,
$\sum_q V_q^2 Im \chi _{cf} (q, \omega)$, is either independent of
$\omega$ and $T$, or depends on them only through a combination
$\omega/T$. The soft energy scale $T_{cf}^*$ is defined such that
this condition is satisfied at $\omega,T \gg T_{sf}^*$.
At $\omega,T \ll T_{cf}^*$, $Im \chi _{cf} (q, \omega)$ is linear
in $\omega$. As a result, $\left( {1 \over \tau_{tr}} \right)_{cf}$
and $\left( {1 \over \tau_{tr,s}}\right)_{cf} $ are both
quadratic in $T$ at $T \ll T_{cf}^*$. We conclude that,
if only charge fluctuations are important, the spin and electrical
resistivities will always have the same temperature dependence.

The spin fluctuation contribution can be considered in a similar
fashion. The corresponding contributions to the transport and spin
transport scattering rates are given as follows,

\begin{eqnarray}
\left( {1 \over \tau_{tr}}\right)_{sf} = &&
{3 \over  A}
\sum_{k, q} \gamma_{kq}^2
B(k, q) J_q^2 Im \chi _{sf} (q, \epsilon_{k+q}-\epsilon_{k})\nonumber\\
\left( {1 \over \tau_{tr,s}} \right)_{sf} =&& {1 \over 3} \left(
{1 \over \tau_{tr}}\right)_{sf} +
{2 \over  A}
 \sum_{k, q} \tilde{\gamma}_{kq}^2 B(k, q) J_q^2 Im \chi _{sf}
(q, \epsilon_{k+q}-\epsilon_{k})
\label{tausf}
\end{eqnarray}
where $\tilde{\gamma}_{kq} =v_{k+q} + v_k$ is the {\it sum} of the group
velocities of the quasiparticle states before and after a scattering event.
Physically, the spin orientation of the quasiparticle is reversed
after a spin-flip scattering, and so is its contribution
to the spin current. Two possibilities need to be considered.
If the spin fluctuation spectrum, $Im \chi _{sf} (q, \omega )$,
is only weakly $q-$dependent, then the
factors $\gamma_{kq}$ and $\tilde{\gamma}_{kq}$ will not lead to
differences in the temperature dependences of
$\left( {1 \over \tau_{tr}}\right)_{sf}$ and $\left( {1 \over
\tau_{tr,s}} \right)_{sf}$. Again, $\left( {1 \over \tau_{tr}}
\right)_{sf}$ and $\left( {1 \over \tau_{tr,s}} \right)_{sf} $ are
linear in $T$ at $T \gg T_{sf}^*$, and quadratic in $T$ at $T \ll T_{sf}^*$,
where $T_{sf}^*$ is the corresponding soft energy scale associated with
the spin fluctuations; the prefactors are in general different, though of
the same order of magnitude.
If $Im \chi _{sf} (q, \omega )$ is sharply peaked
at a particular wavevector $Q$, it is possible that
$\tilde{\gamma}_{kq}$ and ${\gamma}_{kq}$ act as different form
factors leading to different temperature dependences in
$\left( {1 \over \tau_{tr}}\right)_{sf}$ and $\left( {1 \over
\tau_{tr,s}} \right)_{sf}$. Such will be the case if, and only if,
when we expand the $q-$dependence of $\tilde{\gamma}_{kq}$ and
${\gamma}_{kq}$ around $Q$, the leading terms for
$\tilde{\gamma}_{kq}$ and ${\gamma}_{kq}$ have different powers in
$\delta q = (q-Q)$. It is easy to see that, this can occur only if $Q = 0$.
As long as $Q \ne 0$, which can safely be assumed to be
the case for the cuprates\cite{Slichter}, the difference between
$\tilde{\gamma}_{kq}$ and ${\gamma}_{kq}$ can lead to a difference
between $\left( {1 \over \tau_{tr}}\right)_{sf}$ and $\left( {1 \over
\tau_{tr,s}} \right)_{sf}$ only in the overall magnitude, not in the
temperature dependence. Therefore, the spin and electrical resistivities
will {\it both} be linear in $T$ at $T \gg T_{sf}^*$ and quadratic
in $T$ at $T \ll T_{sf}^*$.

We now turn to the situation that the spin fluctuations and
charge fluctuations are both important. If $T_{sf}^*$
and $T_{cf}^*$ are well separated, the spin and electrical resistivities
will have different temperature dependences for temperatures
from ${\rm min}(T_{sf}^*, T_{cf}^*)$ to ${\rm max}(T_{sf}^*, T_{cf}^*)$.
However, the electrical resistivity is linear in $T$ only at
$T \gg {\rm max}(T_{sf}^*, T_{cf}^*)$. In this same temperature range,
the spin resistivity is also linear in $T$.

The discussion of a $T-$linear resistivity in these Fermi-liquid-like
states\cite{Review,Monthoux,mfl} is usually restricted to the level
of semi-classical description, which corresponds to the leading order
terms we have considered so far. Should contributions beyond the leading
order become important, it is not clear how a $T-$linear
resistivity can arise from these Fermi-liquid-based schemes. In this
regard, a particularly relevant possibility involves the transport
of the collective modes themselves. This would occur if, at relevant
length and energy scales, the collective fluctuations,
$\chi_{sf}(q,\omega)$ and/or  $\chi_{cf}(q,\omega)$, in fact
describe some well-defined excitations. In this case,
two additional contributions to the conductivities arise. One describes
the spin conductivity from the spin-wave contribution to the
spin-current. The other corresponds to the fluctuating conductivities
coming from both $\chi_{sf}(q,\omega)$ and $\chi_{cf}(q,\omega)$, as
described by the Aslamasov-Larkin-type diagrams\cite{paracond}. From a
general analysis of the vector vertices,
it can be shown that their contributions to the electrical conductivity
and spin conductivity are in general different. While these additional
contributions might be of relevance to the underdoped cuprates\cite{sw},
for the optimally doped case, given the simple behavior of the observed
thermodynamic properties (the essentially $T-$independent
susceptibility\cite{Review} and specific heat coefficient\cite{Loram})
and an antiferromagnetic correlation length of the order of a lattice
spacing, it is expected that these collective transport should be
negligible. We therefore conclude that, if the $T-$linear resistivity
in the optimally doped cuprates originates from quasiparticle scatterings
off of soft collective modes, the spin resistivity
should also be $T-$linear.

\noindent
\underline{\bf Spin-Charge Separated States: Luttinger Liquid}
Different kinds of spin-charge separation may occur,
and we will illustrate our idea by considering several
examples. First, the Luttinger liquid in 1D\cite{1Dlut}.
Here, the spin and charge excitations propagate with different velocities,
and a complete spin-charge separation is realized.
Linearizing the dispersion around the two Fermi points, and
introducing a boson representation of the fermion fields,
the general interacting spin$-{1 \over 2}$ Fermion model in 1D
can be written as,

\begin{eqnarray}
H_{lut} = &&H_{\rho} + H_{\sigma}+ H_{g_3} + H_{g_1}
\label{hamlut0}
\end{eqnarray}
where

\begin{eqnarray}
H_{\nu} = &&{1 \over 2\pi} v_{\nu} \int dx ~~[ K_{\nu} (\pi \Pi_{\nu})^2
+ {1 \over K_{\nu}} (\partial_x \phi _{\nu})^2]
\label{hamlut1}
\end{eqnarray}
describes the kinetic term for the free charge ($\nu=\rho$)
and spin ($\nu=\sigma$) bosons, $\phi_{\rho}$ and $\phi_{\sigma}$.
Here, $\Pi_{\rho}$ and $\Pi_{\sigma}$ are the corresponding
conjugate momenta. The charge and spin velocities,
$v_{\rho}$ and $v_{\sigma}$, as well as
the charge and spin coupling constants, $K_{\rho}$ and $K_{\sigma}$,
are determined by the forward scattering interactions.
The Umklapp interaction

\begin{eqnarray}
H_{g_3} = && {g_3 \over (2 \pi a )^2} \int dx ~~{\rm cos} (\sqrt{8}
\phi_{\rho} + \delta x )
\label{hamlut2}
\end{eqnarray}
describes two electrons with opposite spins being scattered from one
Fermi point to another.
Here, $a$ is a cutoff parameter, and $\delta$ measures the deviation
from half-filling. The backscattering interaction

\begin{eqnarray}
H_{g_1} = && {g_1 \over (2 \pi a )^2 }\int dx ~~ {\rm cos} (\sqrt{8}
\phi_{\sigma} )
\label{hamlut3}
\end{eqnarray}
describes two electrons, from the opposite Fermi points and with
opposite spins, interchanging branches. The electrical resistivity in
this model has been studied extensively by Giamarchi\cite{Giamarchi},
whose notation we follow closely.

The dissipation of the electrical current, $j = (-e){\sqrt{2} \over \pi}
\partial_t \phi_{\rho}$, is due to the Umklapp term.
Away from half-filling, there exists an energy scale\cite{Giamarchi},
$\Delta^* \sim \delta W$ (where $W \sim v_F/a $ is the characteristic
bandwidth), below which all the Umklapp scatterings are frozen. At
$T \ll \Delta^*$, the electrical resistivity goes to zero
exponentially. At $T \gg \Delta^*$, it has the algebraic form with an
interaction-dependent exponent,

\begin{eqnarray}
\rho \sim {{4 \pi} \over {\omega_p^2}} (\rho_o g_3 )^2 W ({T \over W})^
{4K_{\rho}-3}~~~~~~~~~~~~~~~~~~{\rm for} ~~~~~~~~T \gg \Delta^*
\label{rholut}
\end{eqnarray}

In contrast, the dissipation of the spin current, $j_s = (g\mu_B/2)
{\sqrt{2} \over \pi}
\partial_t \phi_{\sigma}$, comes entirely from the backscattering
term and we find that,

\begin{eqnarray}
\rho_{spin} \sim
{{4 \pi} \over {\omega_{p,s}^2}} (\rho_o g_1 )^2 W ({T \over W})^
{4K_{\sigma}-3}
\label{rhoslut}
\end{eqnarray}
for all temperatures.  $K_{\rho}$ and $K_{\sigma}$ are different for
any non-zero interaction. Within the repulsive Hubbard model,
the exponents for $\rho$ and $\rho_{spin}$ can differ by as large as 2.

\noindent
\underline{\bf Spin-Charge Separated States: Gauge Theory of the
2D $t-J$ Model.} We now consider the gauge theory of the $t-J$
model in two dimensions\cite{Gauge}. This theory describes a state
with spinon-like and holon-like excitations coupled by a
massless transverse gauge field. The presence of this coupling to
the gauge field leads to a situation in between that of a complete
spin-charge separation and that of no spin-charge separation. As we will
see, such an intermediate situation is also reflected in the
relationship between $\rho_{spin}$ and $\rho$.

In terms of the slave-fields $f_{i\sigma}^{\dagger}$ and $b_i^{\dagger}$,
which create singly occupied and empty configurations respectively,
the $t-J$ model is defined by the following Hamiltonian

\begin{eqnarray}
H_{tJ} = -t \sum_{<ij>} (f_{i\sigma}^{\dagger} b_i)
(b_j^{\dagger} f_{j\sigma} )
+ J \sum_{<ij>}\sum_{\sigma \sigma'} (f_{i\sigma}^{\dagger} f_{i\sigma'})
(f_{j\sigma'}^{\dagger} f_{j\sigma})
\label{hamtj}
\end{eqnarray}
with a no-double-occupancy constraint

\begin{eqnarray}
\sum_{\sigma}f_{i\sigma}^{\dagger} f_{i\sigma} + b_i^{\dagger}b_i = 1
\label{constraint}
\end{eqnarray}
at every site $i$. In Eq. (\ref{hamtj}), $<ij>$ labels the nearest neighbors,
$t$ the hopping amplitude, and $J$ the exchange interaction.
The gauge theory description applies when there is a uniform
nearest neighbor RVB order parameter, $\sum_{\sigma}
<f_{i\sigma}^{\dagger}f_{j\sigma}> = \Delta_o$,
and when the boson field is not condensed. The latter ensures
the existence of the massless transverse (unscreened) gauge field,
which formally corresponds to the phase of $\Delta_o$. This gauge field
is coupled to both the bosons and fermions described
by the $b-$ and $f-$ fields respectively. The electrical
current operator is given by $j_x = (-ite)\sum_{i\sigma}
[ (f_{i\sigma}^{\dagger} b_i) (b_{i+\vec{x}}^{\dagger} f_{
{i+\vec{x}}\sigma} )-H.c.]$. The current-current correlation function
can be shown\cite{ioffelarkin} to be

\begin{eqnarray}
\Pi_{jj} = \Pi_{j_fj_f} - (\Pi_{j_fj_f})^2 [\Pi_{j_fj_f} +\Pi_{j_bj_b}]^{-1}
\label{cctj}
\end{eqnarray}
where $\Pi_{j_bj_b}$  and $\Pi_{j_fj_f}$ are the current-current correlation
functions for the boson and fermion currents,
$j_{b}=-e\sum_{k\sigma} v_k^{b} b_{k\sigma}^{\dagger} b_{k\sigma}$
and $j_{f}=-e\sum_{k\sigma} v_k^{f} f_{k\sigma}^{\dagger}
f_{k\sigma}$, where $v_k^f$ and $v_k^b$ are the fermion and boson velocities.
The second term  of Eq. (\ref{cctj}) reflects the screening by the gauge
field. Such a screening enforces the no-double occupancy constraint.
Eq. (\ref{cctj}) leads to an electrical resistivity

\begin{eqnarray}
\rho = \rho_b + \rho_f
\label{rhotj}
\end{eqnarray}
where $\rho_b$ and $\rho_f$ are the boson and fermion resistivities,
respectively, reflecting the scattering of the bosons and fermions
by the gauge field. Using $\rho_f \sim
{{4 \pi} \over {\omega_{p,f}^2}} E_f (T/E_f)^{4/3}$\cite{Lee},
and $\rho_b \sim {{4 \pi} \over {\omega_{p,b}^2}} E_b (T/E_f)$\cite{Gauge},
where $E_f \sim J$, $E_b \sim t$, and ${{\omega_{p,f}^2}/
{\omega_{p,b}^2}} \sim {1 \over \delta}$ with $\delta$ representing
the doping concentration, the resistivity is approximately linear
in temperature.

Using the spin current operator,
$(j_s)_x = (g\mu_B) \sum_{i\sigma} \sigma
[(it/2) (f_{i\sigma}^{\dagger} b_i) (b_{i+\vec{x}}^{\dagger}
f_{{i+\vec{x}}\sigma} )
- (iJ/4) (f_{i\sigma}^{\dagger} f_{i\bar{\sigma}})
(f_{{i+\vec{x}}\bar{\sigma}}^{\dagger} f_{{i+\vec{x}}\sigma})-H.c.] $,
we find that the dominant contribution to the
spin current-current correlation function is given as follows,

\begin{eqnarray}
\Pi_{j_sj_s} = \Pi_{j_{fs}j_{fs}} - (\sum_{\sigma \sigma '}\sigma
\Pi_{j_{fs}^{\sigma}j_{fs}^{\sigma'}})^2 [\Pi_{j_{fs}j_{fs}}
+\Pi_{j_bj_b}]^{-1}
\label{ccstj}
\end{eqnarray}
where $\Pi_{j_{fs}^{\sigma}j_{fs}^{\sigma'}}$ corresponds to
the current-current correlation function for the current operators
$j_{fs}^{\sigma} =(g\mu_B/2) \sum_{k} v_k^f f_{k\sigma}^{\dagger}
f_{k\sigma}$, and $\Pi_{j_{fs}j_{fs}}
=\sum_{\sigma \sigma'}\Pi_{j_{fs}^{\sigma}j_{fs}^{\sigma'}}$.
The second term in Eq. (\ref{ccstj}) again comes from the screening
of the gauge field. However, this term vanishes!
Eq. (\ref{ccstj}) implies that,

\begin{eqnarray}
\rho_{spin} \sim
{{\omega_{p,f}^2}\over{\omega_{p,s,f}^2}}\rho_f
\sim
{{4 \pi} \over {\omega_{p,s,f}^2}} E_f (T/E_f)^{4/3}
\label{rhostj}
\end{eqnarray}
As a result of spin-charge separation, the spin resistivity
is {\it not} linear in temperature despite of a $T-$linear
electrical resistivity.

To summarize, we have demonstrated that it is possible to diagnose
spin-charge separation in the optimally doped cuprates
through a comparison of the spin and electrical conductivities.
If the measured inverse spin diffusion constant turns out to
be not linear in $T$ in, for example, the optimally doped
YBCO or LSCO for which the electrical resistivity is known to be
$T-$linear and the uniform static spin susceptibility essentially
$T-$independent, it would provide a direct evidence for spin-charge
separation in the cuprates. Finally, our analysis also suggests that
measuring the spin-diffusion constant in the quasi-one-dimensional
materials can help clarify the spin-charge separation theoretically
expected in these systems.

I am grateful to L. Greene and N. Hass for extensive discussions
about the spin-injection experiment in YBCO, to S.-W. Cheong,
A. Chubukov, G. Kotliar, A. J. Millis, D. Pines, T. M. Rice,
E. Shimshoni, S. Sondhi and, especially, A. J. Leggett
for helpful discussions at various stages of this work, and
to the theory group at the AT\&T Bell Labs, where part of
this work was carried out, for their hospitality. This work has
been supported by NSF DMR 912-0000.

%\end{narrowtext}
%\end{twocolumn}

\begin{references}
\bibitem{notetensor} We restrict here to the component diagonal both
in real space and in spin space.
\end{references}
\end{document}